\documentclass[10pt,conference]{IEEEtran}

\usepackage{cite}
\usepackage{amsmath,amssymb,amsfonts}
\usepackage{algorithmic}
\usepackage{algorithm}
\usepackage{graphicx}
\usepackage{textcomp}
\usepackage{xcolor}
\usepackage[hyphens]{url}

\def\BibTeX{{\rm B\kern-.05em{\sc i\kern-.025em b}\kern-.08em
    T\kern-.1667em\lower.7ex\hbox{E}\kern-.125emX}}

\newif\ifblind\blindfalse

\begin{document}

\title{Protecting CPU AI On Edge TEEs: \\WebAssembly's Promise and Practical Challenges}

\author{
\ifblind
\IEEEauthorblockN{
Anonymous authors
}
\else
\IEEEauthorblockN{
Friedrich Vandenberghe\IEEEauthorrefmark{1},
Lachlan J. Gunn\IEEEauthorrefmark{2},
Bruno Volckaert\IEEEauthorrefmark{1},
and Merlijn Sebrechts\IEEEauthorrefmark{1}
}\\
\IEEEauthorblockA{
\begin{minipage}{.5\textwidth}
\centering
\IEEEauthorrefmark{1}IDLab, Department of Information Technology\\
Ghent University - imec, Belgium\\
friedrich.vandenberghe@ugent.be
\end{minipage}%
\begin{minipage}{.5\textwidth}
\centering
\IEEEauthorrefmark{2} Aalto University, Finland
\end{minipage}%
}
\fi
}

\maketitle

\begin{abstract}
AI models on edge hardware contain important intellectual property (IP), but an adversary can steal it when they achieve root access. Trusted Execution Environments (TEE) like Arm TrustZone protect against these Operating System (OS) level attacks. However, they are challenging to use. More precisely, it is difficult to run unaltered applications inside a TEE. This work presents a solution that allows the execution of unaltered AI models, compiled to WebAssembly, on the WebAssembly Micro Runtime (WAMR) in OP-TEE for Arm TrustZone. Additionally, this work provides an AI model distributor that encrypts the WebAssembly binary and places the encryption key in one of the device's fuses. This way, only the WAMR Trusted Application (TA) in OP-TEE can decrypt and execute the AI model. A thorough evaluation of our solution shows that it incurs an additional overhead of 22\% in comparison to an application manually ported to OP-TEE, while also facilitating the execution of unaltered AI models with an additional inference latency of 6\% without significant porting effort. Overall, there is a pressing need to safeguard the IP of AI models and this work shows that there is a real promise in WebAssembly, but there remain some practical challenges.
\end{abstract}

\begin{IEEEkeywords}
WebAssembly, Confidential Computing, Arm TrustZone, Computer Vision, Secure Edge.
\end{IEEEkeywords}
    
\section{Introduction}

AI models on commercial, off-the-shelf edge hardware can contain important intellectual property (IP), but its software alone is limited in its ability to protect this IP from being extracted by an attacker. An adversary that achieves root access can extract all data from the device, including AI model weights.  One way to protect the AI models is via a hardware-protected Trusted Execution Environment (TEE), but this brings its own challenges; as a TEE like Arm TrustZone~\cite{arm_trustzone,arm_trustzone_2} provides a very constrained environment for developers~\cite{libutils}, providing an experience more akin to an embedded platform. Thus, developers cannot simply recompile existing applications to run inside TEE. Most environments only provide a limited variant of the C standard library, and lack support for Linux system calls. As a result, executing AI models within a TEE requires significant porting effort and specialized expertise, and must be repeated for each TEE platform. There is thus a pressing need for a mechanism to run existing AI software inside edge TEEs such as Arm TrustZone without significant porting effort.

\begin{figure}
   \includegraphics[width=\linewidth]{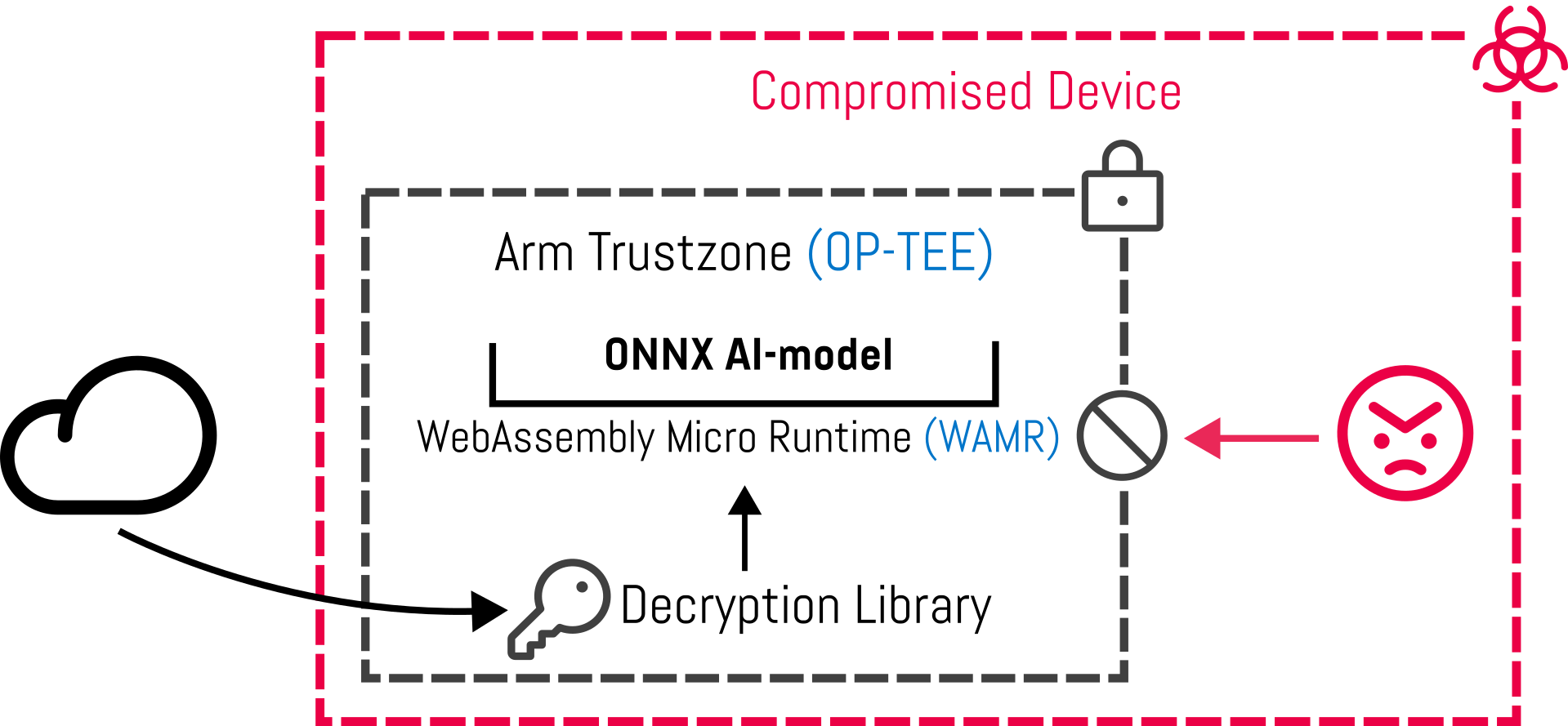} 
   \centering
   \caption{
   The WebAssembly Micro Runtime (WAMR) allows unaltered AI applications to run inside an edge Trusted Execution Environment (TEE) such as OP-TEE for Arm TrustZone. \textbf{Takeaway:} The TEE ensures that an adversary on a compromised Operating System (OS) \textbf{cannot exfiltrate the AI model}, while \textbf{WebAssembly negates the porting effort} required to run the AI model inside the TEE.}
   \label{fig:overview}
\end{figure}

While there exist some efforts that aim to already alleviate this porting effort, they are limited in their language support or their support for highly constrained devices. One is TREO~\cite{treo}, which runs an application written in the C programming language directly inside of the Arm TrustZone Secure World with few or no modifications. A downside to this is that it supports only applications written in C.
PARTEE~\cite{partee} is a recent effort that provides container-like partitions built on Arm TrustZone, but that is currently not yet open-source nor is it generally available. Alternatively, WebAssembly is a promising technology that also offers container-like environments, for example WaTZ~\cite{watz} shows that it is possible to execute existing general applications inside of Arm TrustZone via the integration of a Wasm virtual machine.
Yet, this integration did not deal with the execution of unaltered AI models. Furthermore, this integration also has some lacking traits that are blocking for easily executing AI models. For example, WaTZ uses a remote attestation protocol to release keys and confidential information to the TEE. This requires a connection with a trustworthy second party to initiate the AI model, which is incompatible with some use-cases where devices boot up without an internet connection. 
These existing efforts are examined in-depth in section~\ref{sec:related}.

To answer the pressing need to safeguard the IP of AI models, we present a WebAssembly-based AI platform and model distribution methodology integrated with Arm TrustZone.

This platform makes use of the WebAssembly Micro Runtime~\cite{wamr} and is then evaluated using an NVIDIA Jetson AGX Orin. Our integration works fully offline and offers support for AI models compiled to the Open Neural Network Exchange (ONNX)~\cite{onnx} format. This way, we can easily port existing applications to run inside the TEE, thereby fulfilling the promise of protecting important AI model IP against adversaries with root access. Fig.~\ref{fig:overview} illustrates a conceptual overview. Some practical challenges do endure, e.g. opportunities remain in the current WebAssembly ecosystem to improve AI model support and the strict memory handling imposed by Arm TrustZone remain. 


In this paper, we aim to provide an answer to the following \textsc{\textbf{research questions}}:
\begin{enumerate}
    \item[RQ1)] How does WebAssembly perform as a way to recompile existing AI models for Arm TrustZone without any modifications?
    \item[RQ2)] What is the performance overhead of the execution of Web\-As\-sem\-bly-applications inside Arm TrustZone in comparison with native code?
    \item[RQ3)] How have new developments in the WebAssembly Micro Runtime impacted the performance of WebAssembly in TEEs?
\end{enumerate}


Moreover, our \textsc{\textbf{contributions}} to the state of the art are:

\begin{enumerate}
    \item A runtime to execute modern WebAssembly-based workloads inside of the Arm TrustZone Secure World.
    \item A demonstration of our solution applied to a computer vision use-case.
    \item Benchmarks measuring performance overhead relative to TEE-specific native code.
    \item A discussion on the limitations and future of Web\-As\-sem\-bly for machine-learning tasks
\end{enumerate}

\section{Background}


The proposed solution supports trusted execution of encrypted WebAssembly code using Arm TrustZone. This section concisely explains the underlying technologies of the solution.

\subsection{Arm TrustZone}


The Arm TrustZone technology, introduced in Armv6, adds an orthogonal set of privileges which divide the system into `Normal World' and `Secure World' modes, activated respectively when executing code from the Rich Execution Environment (REE) or the Trusted Execution Environment (TEE). Typically, the REE contains a Linux OS, and the TEE uses a specialized TEE OS with its own Application Programming Interface (API).
Application software inside the Secure World, called a Trusted Application (TA), has access to the secure physical address space, which cannot be accessed by the Normal World. The processor switches between executing code from the Normal World and the Secure World in a time-sliced fashion via the Secure Monitor Call (SMC) instruction.

Another variant of TrustZone, \emph{TrustZone-M}, is used in ARM microcontrollers~\cite{arm_m_profile}.  TrustZone-M functions quite differently to the TrustZone extensions in application processors, and will not be considered in this paper. 

\subsection{OP-TEE}

OP-TEE~\cite{optee} is an open-source TEE OS that implements an API for the communication between TEE and TA. This API is defined by GlobalPlatform (GP)~\cite{gp}, an organisation that works across industries to publish specifications which facilitate the secure and interoperable deployment of embedded applications on secure chip technology. Besides the API implementation, OP-TEE also provides a TEE Operating System (OS) and a Linux kernel TEE framework and driver to facilitate communication between the TEE and REE. A part of the OP-TEE core is an interface to other Trusted Applications and non-secure client applications. This interface is provided by so-called Pseudo Trusted Applications (PTAs)~\cite{optee-pta}. 
Using all these components, OP-TEE facilitates the transition from the Normal World, via a Secure Monitor and TEE OS, to the TA. 


Due to the nature of how OP-TEE is designed, there are some aspects that heavily limit application development and which make it impossible to directly execute unaltered applications. Firstly, there is no C standard library; TAs are instead programmed against the GlobalPlatform API. Secondly, it  is impossible to spawn a new thread inside a TA~\cite{optee-mt}, and thus there is no support for multithreading inside the Secure World. Thirdly, OP-TEE only provides first-class support for the C and C++ programming languages, and experimental support for Rust. Lastly, for an application to execute inside the Secure World, it needs to adhere to the GP API. 


\subsection{WebAssembly}

WebAssembly (Wasm)~\cite{wasm} is a binary instruction format for a stack-based virtual machine. Wasm is designed as a portable compilation target for programming languages, enabling deployment on the web for client and server applications. Currently, over seventy programming languages can target Wasm~\cite{wasm-langs}. Outside the web, WebAssembly System Interfaces (WASI)~\cite{wasi} extend Wasm with interfaces that grant capabilities to system resources, e.g. network~\cite{wasi-http}, file system~\cite{wasi-file}, input/output~\cite{wasi-io}, and machine learning~\cite{wasi-nn}. WASI itself is not yet fully mature, but consists of three preview versions: the first preview provides traditional C-like interfaces, preview two replaces these interfaces with the typed memory-safe Application Binary Interface from the WebAssembly Component Model, and the third preview enhances the Component Model with capabilities for concurrency and asynchronous execution. In essence, compiling to Wasm equates to compiling to either core Wasm or one of the three WASI previews.


\subsection{ONNX Runtime}

The AI ecosystem defines a myriad of formats to store AI models. Of these, the Open Neural Network Exchange (ONNX) format~\cite{onnx} distinguishes itself by being interoperable across framework. ONNX achieves this interoperability by defining a common set of building blocks for machine learning and deep learning models, and a common file format to enable developers to use models across the ecosystem. To execute ONNX AI models, developers simply have to pick an ONNX-capable inference runtime. Alternatively, developers can use the \texttt{onnx2c}~\cite{onnx2c} compiler to compile the model in ONNX format to C code designed for inference on microcontrollers. The generated code exemplifies this by requiring only the standard C maths library and allocating the buffers at compile-time. \texttt{onnx2c} currently only implements half of the available ONNX operations, and does not support backpropagation.

\subsection{WebAssembly Micro Runtime}

To execute a Wasm binary, a runtime is needed to translate Wasm instructions to native ones. The lightweight WebAssembly Micro Runtime (WAMR) virtual machine is the de facto standard runtime for edge devices. From the four possible Wasm targets, WAMR has support for core Wasm and the first preview of WASI. 


In WAMR version 2.0.0 the memory model got changed to one where \texttt{mmap} is now used to allocate the Wasm linear memory~\cite{wamr-3052} instead of allocating memory from the shared pool via \texttt{malloc}. Inside OP-TEE, \texttt{mmap} is provided as an extension to the GlobalPlatform API. This extension means that calls to \texttt{mmap} are placed on an event bus that is handled by the OP-TEE core process. The memory is then allocated in the address space outside the one carved out for the TA. In conclusion, memory allocated via \texttt{mmap} is effectively taking away available memory for other TAs, and thus this allocation is doing a Denial of Service on any other TA.

\subsection{Secure Boot}

Secure Boot~\cite{jetson_boot} prevents execution of unauthorized code on the device, using a chain of trust in which the software implementing each stage of the boot process authenticates the software of the subsequent stage before allowing it to execute. The root-of-trust is an on-die read-only memory (ROM) that uses asymmetric-key encryption keys stored in write-once fuses. Storing these keys in fuses, ensures that they cannot be altered when the device is subsequently flashed.



\section{Related Work}\label{sec:related}


Although the Arm TrustZone technology has been available since 2002 with the release of Armv6, it remains challenging to run unaltered applications inside it. Developers cannot recompile existing applications for Arm TrustZone, instead they have to perform significant porting effort. Several studies have explored ways to alleviate this porting effort, which are discussed below.

Mu Yuchen developed Transparent ELF Rehosting for OP-TEE (TREO)~\cite{treo}, which makes it possible to run an original C program directly inside OP-TEE without modification or with only minimal changes. For this, the C program is compiled into an Executable and Linkable Format (ELF) file under a normal Linux environment. However, TREO only provides support for the unaltered execution of applications originally written in C. Furthermore, TREO only supports version 4.2 of OP-TEE, which was released in April 2024.


Habeeb et al. developed PARTEE~\cite{partee}, a firmware platform that is focused on trustworthy execution of AI. Via container-like partitions built on TEEs it enables unmodified Linux applications to run securely. For this, PARTEE provides an asynchronous I/O, built upon Linux's \texttt{io\_uring} features, interface to TEE applications to enable them to request system calls from the host Operating System (OS). These calls are encrypted using standard encryption techniques, like TLS. Furthermore, PARTEE provides a library OS layer that sorts system calls and I/O into either to be handled by PARTEE directly or to be handled by the Normal World OS. While it seems as if PARTEE is promising, there is little information available about the project since the initial white paper in February 2026, nor is it open source or generally available.

Ménétrey et al. developed WaTZ~\cite{watz}, an efficient and secure runtime for trusted execution of Wasm code in Arm TrustZone. Additionally, they developed a lightweight remote attestation system optimised for Wasm applications running in Arm TrustZone. Under the hood, WaTZ makes use of the WAMR. This is a promising approach on which we build in this. However, since the publication of WaTZ in 2022, it has not kept up with developments in the Wasm ecosystem. As a result, the runtime misses a great deal of updates and security features available in current versions of WAMR. Examples of these are 64-bit memory and a garbage collector. These features enable support for more programming languages, and more capabilities. Moreover, WaTZ only supports Arm TrustZone on the MCIMX8M-EVK board.

Zheng et al.~\cite{watz-blockchain} extended WaTZ with blockchain smart contract capabilities, implementing a decentralized identity (DID) smart contract that supports creating DID documents and verifying encrypted verifiable credentials (VCs). Under the hood, they updated WAMR to version 2.2.0 and support Arm TrustZone on QEMU. This shows the power of Wasm to facilitate a broad range of use-cases. Nevertheless, their approach does not address the challenge of protecting the intellectual property of edge-based AI.

\section{Threat Modelling}
    


This section outlines the security model under which the solution is analysed.

We consider a system consisting of a trusted AI model distributor and an untrusted device that contains a secure confidential computing environment. This environment provides memory confidentiality. The boot process of the device is secured against the execution of unauthorized boot codes.

\subsection{Adversary Model}

We consider an adversary that, once the provisioning phase is complete, fully controls the device's REE, as well as any interfaces between the TEE, the REE, and the outside world, including those with the distributor, operating system, and any connections from the operating system with the TEE. The attacker abuses these privileges with the intent to steal the intellectual property (IP) of the AI model.  The attacker cannot compromise the AI Model Distributor itself, nor can it eavesdrop on or interfere with the communication between the AI Model Distributor and the device prior to or during the provisioning phase of the device.



\subsection{Assumptions}

We assume that the secure confidential computing environment correctly enforces memory confidentiality and that the security primitives it relies on are secure. This includes the boundary between the confidential computing environment and the operating system, such that there are no vulnerabilities that enable an adversary to transition to the Secure World without mediation by the Secure Monitor. We assume that all software in the TEE is secure, except for the Wasm modules containing the application software.  We also assume the Secure Boot functionality of the device does not allow any malicious or otherwise unintended software to run in the TEE.  We also assume that the connection between the distributor and the device only allows the device to receive the encrypted AI model.

\subsection{Out-of-scope Threats}

We deem some attacks out-of-scope. These attacks include side-channel attacks, fault-injection attacks, hardware supply-chain attacks, and other physical attacks that require invasive access to hardware, and cryptanalytic attacks by a quantum computer. Examples of such attacks are TEE.fail~\cite{tee-fail}, Battering RAM~\cite{batteringramsp26}, and Differential Power Analysis~\cite{dpa}.

\subsection{Security Objectives}

The following security objectives are described here and evaluated.

\begin{enumerate}
    \item \textbf{Model confidentiality:} The confidentiality of the model remains protected during commissioning, transfer from the normal into the Secure World, and its execution.
    \item \textbf{Platform confidentiality:} For any adversary that controls the device after provisioning, the confidentiality of any application within the Secure World is preserved.
\end{enumerate}

\section{Architecture}

\begin{figure}
   \includegraphics[width=\linewidth]{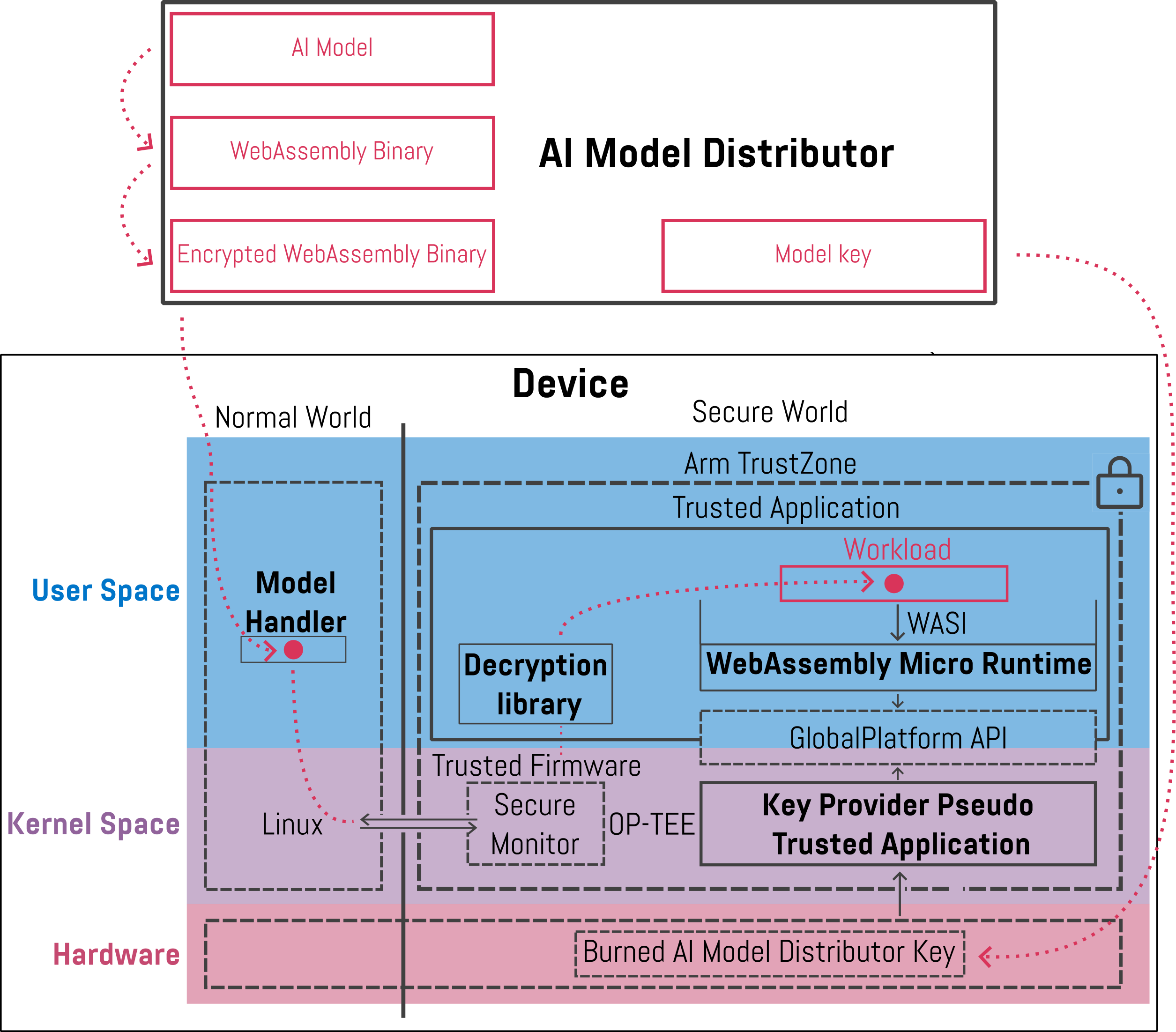} 
   \centering
   \caption{In our architecture, a trusted AI model provider hosts an encrypted Wasm binary, and their key is burned to the hardware of the device. This encrypted binary is passed from the Normal World, via the secure monitor, to the Secure World. In the Secure World, the Wasm binary's signature is verified with the public key of the AI model distributor. Afterwards, the binary is decrypted using the provider's key and then executed in WAMR. 
   \textbf{Takeaway:} The Normal World can never see the contents of the encrypted Wasm binary, and the execution is fully offline.}
   \label{fig:architecture}
\end{figure}

\begin{figure*}[ht]
    \includegraphics[width=\textwidth]{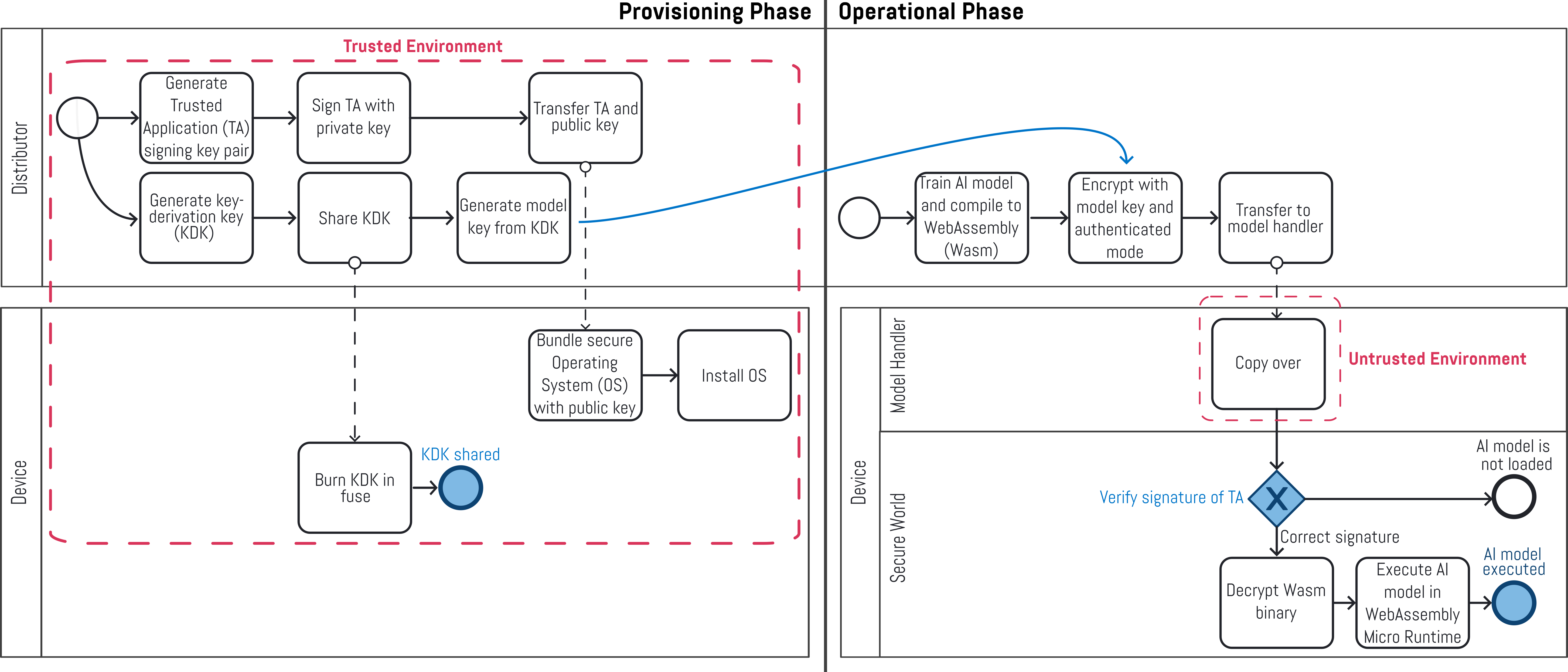}
    \centering
    \caption{Process diagram illustrating the two flows within the proposed architecture that ensure the secure execution of AI models. 
    \textbf{Takeaway:} Thanks to the steps performed in the provisioning phase, an application in the Normal World cannot alter the AI model.}
    \label{fig:flow}
\end{figure*}

Figures~\ref{fig:architecture} and~\ref{fig:flow} illustrate, respectively, the components in our proposed architecture and a detailed process diagram. The AI model distributor transfers an encrypted Wasm binary to the Normal World of the device. In the Normal World, the application transfers the binary to its corresponding Trusted Application (TA) in the Secure World via the Secure Monitor. Because the TA resides in the Secure World, it can access both the public key and the symmetric encryption key of the AI model distributor. With these keys, the trusted application respectively verifies the signature of the Wasm binary, decrypts it, and runs it in the WAMR. 

\subsection{Provisioning}

In the provisioning phase, the AI model distributor generates a TA-signing key pair, and embeds the public part into the TEE OS image, then signs it with the device's secure boot key and loads it onto the device, along with the TA. It then generates a Key-Derivation Key (KDK) and burns it into the device's protected fuses, where it is inaccessible except by the TEE. The use of a KDK rather than encryption and authentication model keys directly allows for crypto system agility, even when the key is stored in fuse memory.  



\subsection{AI model Distributor}



The AI model distributor is a trusted party that trains, signs and encrypts an AI model with a model key that they generated. This architecture proposes two methods to distribute the model to the device. The distributor could provision this model to the device via a secure network connection. During the setup of the device, the device actor will then use this connection to pull in the model. Alternatively, the distributor could place the model on a storage device. This storage device is then placed inside the device.

\subsection{Model Handler}


The model handler is an application that is responsible for maintaining a secure connection with the AI model distributor. This handler also contains an application in the Normal World that is connected with the TA. Upon execution of the application, it will use this connection to place the encrypted AI model in a shared buffer, such that the model is accessible in the Secure World.

\subsection{Extended WebAssembly Runtime}


The extended Wasm runtime TA, based on that of WaTZ~\cite{watz}, contains the WAMR runtime extended with Arm TrustZone platform support and a module that can decrypt the Wasm binary using the KDK. When executed, the TA takes the encrypted AI model binary from the model handler, then verifies the signature, authenticates, decrypts and executes it using WAMR. 



\section{Implementation}



To validate our proposed architecture, we developed a proof-of-concept implementation that executes an AI model. Our primary objectives were to confirm the feasibility of the approach and identify potential opportunities or shortcomings in the underlying architecture and security methodologies.

\subsection{AI model}

For the AI model in the proof-of-concept implementation, we make use of a computer vision model provided by ONNX~\cite{onnx_mnist} that was trained on the MNIST dataset~\cite{mnist} to perform handwritten digit recognition. We convert this model with \texttt{onnx2c}~\cite{onnx2c} from the ONNX format to a C file, allowing it to be compiled to Wasm with the Clang compiler. Finally, we encrypt and sign the binary and transmit it to the device.

\subsection{WebAssembly Runtime}

In our proof-of-concept implementation, the TEE OS is OP-TEE running in the Arm TrustZone Secure World. The TA is forked from WaTZ, modified  primarily around the new Ahead-of-Time (AOT) Application Binary Interface (ABI) introduced in version 2.0.0 of WAMR. In addition, we extended WaTZ to support the NVIDIA Jetson AGX Orin, and support OP-TEE version 4.8 in the QEMU virtual machine.


\subsection{Encryption of the WebAssembly binary}

%

The WebAssembly binary is not directly encrypted with the symmetric encryption key burned into the device fuses, but instead derived from the KDK, we encrypt our Wasm binary with an AES key that is derived via the NIST-SP 800-108 Key Derivation Function (KDF) in counter mode~\cite[§4.1]{nist-kdf} from an AES KDK. We instantiate the pseudorandom function (PRF) with the  AES Cipher-based Message Authentication Code (CMAC), as this is the only option supported by the device's hardware accelerator. The label is a string that identifies the purpose for the derived keying material, hardcoded as \texttt{ekb} and the context is hardcoded as \texttt{encryption}.  This key is generated by the Distributor during the provisioning phase, and then stored for later use during the operational phase.

The encryption itself is 128-bit AES-GCM; this authenticated encryption mode allows us to avoid separately computing an authentication tag.

\section{Performance Evaluation}


This section evaluates the end-to-end performance of our proposed architecture, which consists of startup and runtime latency. We measure both latencies independently. For the runtime latency, we measure the inference latency to infer an image of Grace Hopper on a MobileNetV2~\cite{mobilenetv2} ONNX model. Then, we execute the PolyBench suite~\cite{polybench} to identify any runtime latency implications compared to native execution. All benchmark results are available~\cite{idlab-watz} on GitHub.


\subsection{Test Setup}

To measure the startup latency, we created ten Wasm binaries from which their AOT-compiled versions range from 1 to 10 MiB. Afterwards, we encrypted each individually. These binaries are then executed 500 times via WAMR inside OP-TEE.
To assess the runtime latency of the WAMR inside OP-TEE, we perform 500 iterations of the whole PolyBench suite compiled as a single application. To assess the inference latency, we perform 500 executions of a MobileNetV2 model inferring an image of Grace Hopper. Both applications are then executed both on OP-TEE and on Linux, and both via WAMR and native. All executions are pinned to certain processor cores. Given that we solely wish to measure the runtime latency of WAMR across the executions, we make use of the already decrypted Wasm binary and omit the signature verification.
Furthermore, the machine we use is an NVIDIA Jetson AGX Orin Developer Kit~\cite{jetson} that runs Ubuntu 22.04 as its Rich Execution Environment (REE), and OP-TEE 4.2.0 as its Trusted Execution Environment (TEE). More precisely, the TEE is a fork of OP-TEE 4.2.0 provided by NVIDIA that adds support for their platforms~\cite{jetson_optee}.

\subsection{Startup latency}

\begin{figure}[htbp]
   \includegraphics[width=\linewidth]{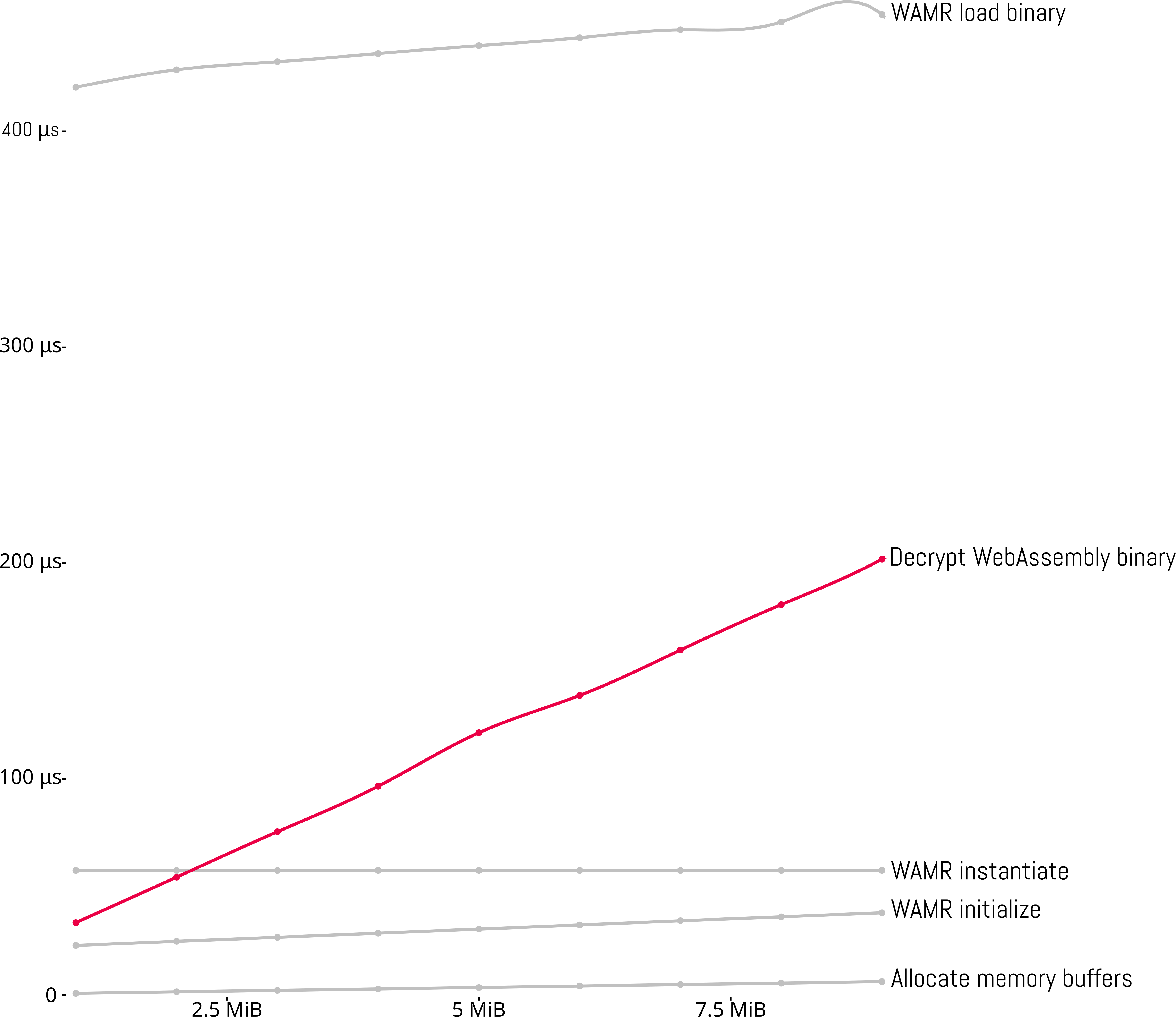} 
   \centering
   \caption{Execution times of each part of the startup procedure inside the TA  with an AOT compiled Wasm binary, going from 1 to 10 MiB. \textbf{Takeaway}: The decryption time scales linearly with the binary size.}
   \label{fig:startup}
\end{figure}

Fig.~\ref{fig:startup} shows the execution time of various parts of the startup procedure across the different sizes. As the size of the encrypted binary increases, the decryption time increases linearly. All other parts of the startup procedure inside the TA remain fairly constant across binary sizes.
\subsection{Inference Latency}

Table~\ref{tab:infer_latency} illustrates the inference latency across three out of the four environments. The OP-TEE native environment is missing due to the size of the MobileNetV2 ONNX model (13.59 MiB) and the then generated C code (14.2 MiB) by \texttt{onnx2c}, which makes the resulting TA too large (14.9 MiB) for OP-TEE to load into the translation tables~\cite{optee-ta-size}. The incurred slowdown on the inference latency is relatively small. Specifically, switching from the REE WAMR environment to TEE WAMR results in a mean slowdown of $21.7 \mu s$ ($+ 5.61\%$).

\begin{table}[htbp]
    \caption{Inference latency of MobileNetV2.} 
    \centering
    \begin{tabular}{cc|ccc}
        \multicolumn{2}{c}{\textbf{Environment}} & \textbf{Min ($\mu s$)} & \textbf{Mean ($\mu s$)} & \textbf{Max ($\mu s$)} \\
        OP-TEE & WebAssembly & 408.36 & 408.42 & 408.55 \\
        Linux & WebAssembly & 386.38 & 386.72 & 386.86 \\
        Linux & Native & 201.13 & 201.18 & 201.22 \\ 
    \end{tabular}
    \label{tab:infer_latency}
\end{table}

\subsection{WebAssembly vs Native Code}



\begin{figure}[htbp]
   \includegraphics[width=\linewidth]{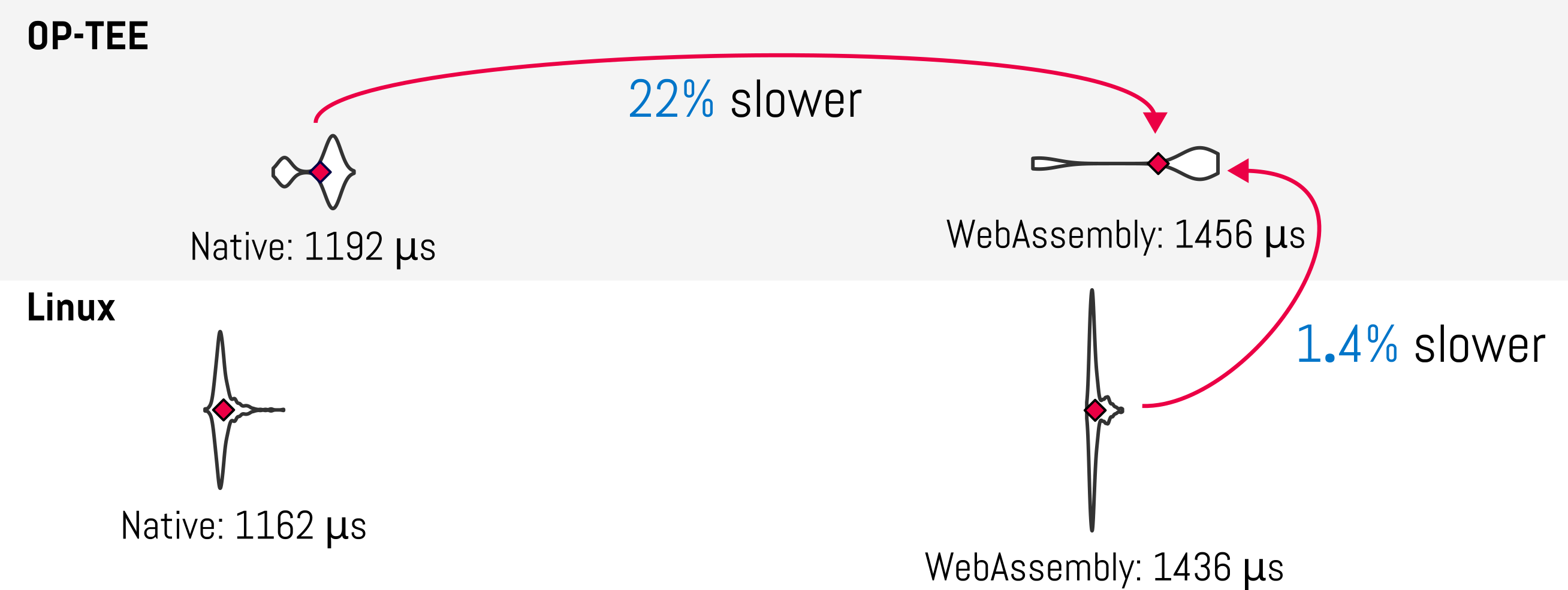} 
   \centering
   \caption{Execution times of running the PolyBench suite as a single application. Red diamond is the average. \textbf{Takeaway}: Main slowdown is because of WAMR, not because of OP-TEE.}
   \label{fig:results_overview}
\end{figure}

Fig.~\ref{fig:results_overview} and table~\ref{tab:results_overview} illustrate the mean execution time for the four environments. The performance overhead is relatively small. Specifically, the greatest overhead does not originate from switching from the REE to the TEE ($+2.62\%$), but from making use of WAMR ($+23.6\%$). 
Furthermore, going from a native execution inside the TEE to using WAMR results in a slower execution of $263.67\mu s$ ($+22.11\%$), whereas the switch from REE WAMR to TEE WAMR only introduces a slowdown of $19.85 \mu s$ ($+1.38\%$).

\begin{table}[htbp]
    \caption{Execution times of running the PolyBench suite as a single application.} 
    \centering
    \begin{tabular}{cc|cc}
        \multicolumn{2}{c}{\textbf{Environment}} & \textbf{Mean ($\mu s$)} & \textbf{Standard deviation ($\mu s$)} \\
        OP-TEE & WebAssembly & 1456.18 & 22.20 \\
        OP-TEE &  Native & 1192.51 & 6.85 \\
        Linux & WebAssembly & 1436.33 & 2.04 \\
        Linux & Native & 1162.10 & 2.93 \\ 
    \end{tabular}
    \label{tab:results_overview}
\end{table}

\subsection{Comparison with Original WaTZ}


A performance evaluation of our solution in comparison with the original WaTZ shows that our solution is slightly faster. For this evaluation, we make use of the same test setup as the other tests, and thus we run the baseline WaTZ on the same hardware.

The WAMR version that our solution extends upon, released 28th of July 2025, is 0.7\% faster than the WAMR version used in WaTZ, dating back to 2021. 
This is not a significant speed-up. The most likely causes are a mix of performance-impacting security enhancements, performance improvements, and a performance-degrading heavier use of the \texttt{mmap} system call.


\section{Threat Evaluation}

This section evaluates the design of the architecture against the three security objectives, considering concrete adversarial actions within the security model. 


\subsection{Platform confidentiality}


An adversary that has root access to the device could attempt to compromise the confidentiality of a Trusted Application (TA), but this is blocked by the security guarantees of Arm TrustZone and OP-TEE. What follows is a listing of how an adversary could attempt to compromise a TA, and how this is thwarted by either Arm TrustZone or OP-TEE.
Firstly, the adversary could try to read or modify the memory of the TA inside the Secure World. During the TA's execution, Arm TrustZone protects the memory of the TA. Furthermore, OP-TEE automatically encrypts all data, including applications, that reside in the Secure World. 
Secondly, the adversary could try to load a custom-designed TA to harm our solution's TA, or to exfiltrate any secrets in the Secure World. This is blocked by OP-TEE. Upon loading of a TA in the Secure World, the OP-TEE OS verifies the signature of the application against an RSA public key bundled with the OP-TEE OS. The private key is not present in OP-TEE; instead, only the trusted AI model distributor has access to this key. During provisioning, the distributor signs our solution offline with their private key. This way, the adversary cannot forge a correct signature for their custom-designed TA.


\subsection{Model confidentiality}

An adversary trying to exfiltrate the AI model, will need to find a way to access and decrypt it. Although it is easy for an adversary to gain access to the encrypted AI model, the adversary cannot access the decryption key.
In our solution, the adversary could gain access to the encrypted AI model when the model handler receives the model from the distributor, and when the handler copies the model to the Secure World. 
To decrypt the AI model, the adversary must obtain the AI model distributor's key. Inside the Secure World, this key is derived with the NIST-SP 800-108 KDF from an AES key that is only readable from within the Secure World.  The use of authenticated encryption prevents an attacker from learning anything about the model by chosen-ciphertext attacks; otherwise, the attacker might obtain a copy of the model weights by flipping bits in the ciphertext and observing the effect on the model outputs.
The encryption of the AI model does not block against denial-of-service attacks while the model is stored on disk, but it does negate attempts to steal the IP of the AI model. The OP-TEE OS hinders alterations during the execution of the AI model by verifying the TA's signature, as this disallows the adversary to load a custom-designed TA.

 



\section{Discussion \& Limitations}


Via an extended WebAssembly Micro Runtime (WAMR) we provide a compatible way to run unaltered applications, and AI models, inside the OP-TEE Trusted Execution Environment (TEE) for Arm TrustZone. This way we can bypass most of the design limitations set forth by OP-TEE with an overhead of 22\%. Only the no multithreading and the memory limits remain a limitation. 
This overhead is a decrease of 0.7\% when compared to WaTZ and is mainly due to updates and WAMR's use of \texttt{mmap}. For AI models, our solution shows a 6\% slower inference latency when running in OP-TEE. Nevertheless, using WAMR our solution can bypass three limitations set forth by OP-TEE: the WebAssembly System Interface (WASI) provides C standard library capabilities, over seventy programming languages can target WebAssembly (Wasm), and an application does not need to adhere to a specific API. Each of these facilitate the execution of unaltered applications inside Arm TrustZone.

\subsection{WebAssembly System Interface}
In theory, the bypass provided by WAMR allows us to easily execute unaltered AI models. But in practice, we are gravely restricted by the limited support for current workloads, especially AI, in the WASI ecosystem. Inside the WASI ecosystem, there is the interface for Machine Learning (ML), called \texttt{wasi-nn}. This interface allows for ML interference workloads inside Wasm, but is by design very different from how ONNX operates. Instead of defining a common set of building blocks for machine learning and deep learning models, \texttt{wasi-nn} is centred around loading a model and a signifier for the ML framework's encoding. This is because \texttt{wasi-nn} is heavily modelled after the Web Neural Network (WebNN) interface~\cite{webnn}, which was unstable and lacking in functionality during the inception of the WASI interface. Nowadays, WebNN is stable enough to be available as an experimental feature in Chromium-based browsers~\cite{ciu-webnn}. Yet, \texttt{wasi-nn} has not been updated to reflect WebNN's current capabilities. If \texttt{wasi-nn} provided the same API as WebNN, it would be possible to run unaltered AI models inside Web\-As\-sem\-bly with near-native execution time. While this is not the case, either an AI runtime must be fully executed within Wasm or ported to OP-TEE. To get around this, we currently compile the AI model into an application that can run without an AI runtime.

\subsection{Arm TrustZone}
For an application in Arm TrustZone, it is impossible to know the maximum available amount of memory without also knowing on which device it runs. This limits the portability inherent to WebAssembly, applications must restrict themselves to remain within the memory limits of their chosen device.
Furthermore, our solution relies heavily on memory-management functions not part of the GlobalPlatform API, but provided as extensions by OP-TEE Core. Thus, they allocate memory from the OP-TEE Core process, which is outside the application's reserved memory space.




\section{Conclusion}

This work enables executing unmodified AI models in a manner such that an attacker with root access cannot exfiltrate them, and thus the intellectual property of the AI models remains safe. While this is a feasible approach, this work identifies some compatibility and performance limitations of Wasm, specifically for accelerated AI models.
To answer RQ1, this work demonstrates the feasibility of executing unaltered AI models inside a TEE. Inspired by earlier approaches such as WaTZ, our solution executes AI models using a WebAssembly runtime inside of Arm TrustZone. While this approach safeguards model confidentiality, it is heavily restricted by the limited set of supported ONNX operators and the limited support for machine learning workloads inside WASI. Furthermore, our solution lacks GPU acceleration, which limits the performance of the AI model. Future work will be necessary to broaden the supported set of AI models, which will require switching to an ONNX runtime and advancing the associated WASI neural network proposal.
Performance-wise, our solution with a recent version of WAMR incurs a 22\% slowdown when compared to a native execution inside OP-TEE, which is 0.7\% faster than the 2021 WAMR version used in WaTZ. Therefore, WAMR's performance has not been considerably impacted over time.  Furthermore, to answer RQ2 and RQ3, our solution incurs a 5.61\% slowdown in the inference latency of MobileNetV2 when compared to an execution of WAMR inside OP-TEE.

\ifblind
\else
\section*{Acknowledgement}


This work received funding from IMEC AAA PSX-PSCEC and the European Union under Grant Agreement number 101225859. Views and opinions expressed are however those of the author(s) only and do not necessarily reflect those of the European Union. The European Union cannot be held responsible for them.

\fi

\bibliographystyle{IEEEtran}
\bibliography{references}
\end{document}